\documentclass[twoside,british,a4paper,final,10pt]{iopart}
\usepackage[T1]{fontenc}
\usepackage[utf8]{inputenc}
\usepackage{geometry}
\usepackage{float}
\usepackage{amstext}
\usepackage{graphicx}

\makeatletter

\providecommand{\tabularnewline}{\\}

\usepackage{iopams}
\usepackage{setstack}

\usepackage[numbers, sort&compress]{natbib}

\newcommand{\eqref}[1]{(\ref{#1})}

\makeatother

\usepackage{babel}
\begin{document}

\title[Copper-boron interactions and LID in solar Si]{A first-principles model of copper-boron interactions in Si: implications
for the light-induced degradation of solar Si }

\author{E Wright$^{1}$, J Coutinho$^{1}$, S Öberg$^{2}$, V J B Torres$^{1}$}

\address{$^{1}$Department of Physics and I3N, University of Aveiro, Campus
Santiago, 3810-193 Aveiro, Portugal}

\address{$^{2}$Department of Engineering Sciences and Mathematics, Luleå
University of Technology, SE-97187 Luleå, Sweden}

\ead{wrighteap@ua.pt}
\begin{abstract}
The recent discovery that Cu contamination of Si combined with light
exposure has a significant detrimental impact on carrier life-time
has drawn much concern within the solar-Si community. The effect,
known as the copper-related light-induced degradation (Cu-LID) of
Si solar cells, has been connected to the release of Cu interstitials
within the bulk {[}\emph{Solar Energy Materials \& Solar Cells}, 147:115-126,
2016{]}. In this paper, we describe a comprehensive analysis of the
formation/dissociation process of the CuB pair in Si by means of first-principles
modelling, as well as the interaction of CuB defects with photo-excited
minority carriers. We confirm that the long-range interaction between
the Cu$_{\mathrm{i}}^{+}$ cation and the B$_{\mathrm{s}}^{-}$ anion
has a Coulomb-like behaviour, in line with the trapping-limited diffusivity
of Cu observed by transient ion drift measurements. On the other hand,
the short-range interaction between the d-electrons of Cu and the
excess of negative charge on B$_{\mathrm{s}}^{-}$ produces a repulsive
effect, thereby decreasing the binding energy of the pair when compared
to the ideal point-charge Coulomb model. We also find that metastable
CuB pairs produce acceptor states just below the conduction band minimum,
which arise from the Cu level emptied by the B acceptor. Based on
these results, we argue that photo-generated minority carriers trapped
by the metastable pairs can switch off the Coulomb interaction that
holds the pairs together, enhancing the release of Cu interstitials,
and acting as a catalyst for Cu-LID.
\end{abstract}


\submitto{\JPCM }
\maketitle

\section{Introduction}

Owing to the extremely high diffusivity of copper in Si, the large
majority of Cu contaminants in the bulk precipitate into silicides,
decorate existing extended defects, such as dislocations or grain
boundaries, or out-diffuse towards the surface \cite{ist02}. Additionally,
a small but unavoidable fraction of Cu atoms linger within the bulk,
and these are capable of forming various electrically-active defects,
like interstitial Cu (Cu$_{\mathrm{i}}$), substitutional Cu (Cu$_{\mathrm{s}}$)
\cite{kna02}, combinations of both these defects, \cite{the07,shi09},
as well as complexes containing copper and other species like oxygen,
transition metals and dopants \cite{kna04}. Interest in the physics
and chemistry of Copper was recently renewed, in the context of the
light-induced degradation (Cu-LID) of solar silicon \cite{lin14},
an effect whereby above-bandgap illumination or forward biasing results
in the reduction of minority carrier life-times in solar cells \cite{has03,bot03},
but which can be prevented through the forced out-diffusion of positively-charged
Cu (Cu$_{\mathrm{i}}^{+}$) impurities \cite{bou14}. Notably, the
underlying recombination centre remains unknown \cite{lin16}.

As a point defect, Cu occurs primarily as $\mathrm{Cu_{i}^{+}}$,
except in $\mathrm{n^{+}}$-Si or in the presence of vacancies, where
Cu is found as Cu$_{\mathrm{s}}^{-}$ \cite{bra04}. Precipitation
occurs mostly in the form of amphoteric Cu silicides ($\mathrm{Cu_{3}Si}$
), and is favoured in n-type material over p-type material \cite{ist98a}.
In fact, precipitation in B-doped Si at room temperature occurs only
for interstitial Cu concentrations $[\mathrm{Cu_{i}}]\gtrsim[\mathrm{B_{s}}]+10^{16}\,\mathrm{cm}^{-3}$,
where $[\mathrm{B_{s}}]$ is the concentration of substitutional B
\cite{fli00}. Below this threshold, the favourable reaction paths
for Cu are out-diffusion or pairing with B atoms driven by the Coulomb
attraction between $\mathrm{Cu_{i}^{+}}$ and $\mathrm{B_{s}^{-}}$
ionic charges \cite{pre92,est99}. 

Although CuB pairs are thought to be electrically inactive, their
association/dissociation dynamics limit the Cu-LID rate, at the very
least — the effect slows down with increasing $\mbox{\ensuremath{\mathrm{[B_{s}]}}}$,
speeds up with temperature, and both the concentration of the underlying
\emph{LID-defect} and the activation energy for the rate of degradation
($E_{\mathrm{a,LID}}$) increase with {[}B{]}, as well as with {[}Ga{]}
\cite{lin14}.

The current understanding of the association/dissociation dynamics
of Cu$A$ pairs (where $A$ stands for a group-III dopant species)
is based on a diffusion-limited trapping model of $\mathrm{Cu_{i}^{+}}$.
In the absence of $A\mathrm{_{s}^{-}}$ dopants, the average thermal
energy of copper ions ($\beta$) causes them to diffuse randomly through
the tetrahedral interstitial sites of the Si lattice \cite{web83,ist98}.
In the presence of $A\mathrm{_{s}^{-}}$ dopants, and for $\mathrm{Cu_{i}^{+}}$-$A\mathrm{_{s}^{-}}$
separations for which the attractive potential matches $\beta$, $\mathrm{Cu_{i}^{+}}$
is assumed trapped by $A\mathrm{_{s}^{-}}$, forming a Cu$A$ pair
with zero net charge. This model was employed to explain the observed
reduction in the diffusivity of $\mathrm{Cu_{i}^{+}}$ as a function
of the acceptor concentration ($N_{a}$), with respect to the intrinsic
value ($D_{\mathrm{int}}$). Accordingly, transient ion drift (TID)
measurements in p-Si have shown that the effective diffusivity of
$\mathrm{Cu_{i}^{+}}$ ($D_{\mathrm{eff}}$) decreases for increasing
acceptor concentrations based on the following semi-empirical expression
\cite{pre92,mes96,ist98,hei98}, 
\begin{equation}
D_{\mathrm{eff}}(N_{a},\beta)=\frac{D_{\mathrm{int}}}{1+\mathrm{(C}/\beta)\,N_{a}D_{\mathrm{int}}\tau_{\mathrm{diss}}},\label{eq:effective_diffusivity}
\end{equation}
where $\beta\equiv k_{\mathrm{B}}T$ in eV ($k_{\mathrm{B}}$ is the
Boltzmann constant and $T$ the temperature) and $\mathrm{C}\equiv e/(\varepsilon_{r}\varepsilon_{0})\times10^{2}$
in $\text{eV}\cdot\text{cm}$ ($e$, $\varepsilon_{\mathrm{r}}$ and
$\varepsilon_{0}$ are the elementary electron charge, the relative
permittivity of the bulk and the vacuum permittivity, respectively
- in SI units), and $N_{a}$ and $D_{\mathrm{int}}$ are in $\mathrm{cm^{-3}}$
and $\mathrm{cm^{2}}\text{/s}$, respectively. The dissociation rate
($\tau_{\mathrm{diss}}$) for the Cu$A$ pair and $D_{\mathrm{int}}$
can be obtained experimentally, and fittings to the corresponding
temperature-dependent data yield the dissociation energy of the Cu$A$
pair ($E_{\mathrm{d}}$) and the migration barrier ($E_{\mathrm{m}}$)
of $\mathrm{Cu_{i}^{+}}$ respectively \cite{pre92,ist98}.

The derivation of Eq.~\ref{eq:effective_diffusivity} assumes that
the covalent interaction between $\mathrm{Cu_{i}^{+}}$ and $A_{\mathrm{s}}^{-}$
is negligible, and the binding energy ($E_{\mathrm{b}}$) of Cu$A$
pairs is independent of the species $A$. Assuming that in the ground
state of the pair (i) the Cu and $A$ ions are separated by a distance
equivalent to that of the crystalline Si-Si bond length, \emph{i.e.}
that they are located at nearest neighbouring interstitial and substitutional
sites, respectively, and that (ii) their interaction is well described
by a two-point-charge model mediated by the static dielectric constant
of Si,\footnote{Although the aim of this paper is not to discuss the type of bond
between Cu and the acceptor, the short distance between the Cu and
$A$ atoms in the ground state implies that a description based on
macroscopic electrostatics is highly debatable.} one arrives at a binding energy $E_{\mathrm{b}}=0.52$~eV and a
value of $E_{\mathrm{d}}\simeq E_{\mathrm{m}}+E_{\mathrm{b}}=0.70$~eV,
where $E_{\mathrm{m}}=0.18\pm0.01\ \text{eV}$ has been determined
experimentally \cite{ist98}. While this produces good agreement with
the experimental values of $E_{\mathrm{d}}$ for CuAl, CuGa, and CuIn
pairs, $0.70\pm0.02\ \text{eV}$, $0.71\pm0.02\ \text{eV}$ and $0.69\pm0.02\ \text{eV}$
respectively, the same is not true for CuB pairs, where $E_{\mathrm{d}}=0.61\pm0.02\ \text{eV}$
\cite{pre92,ist02}. An attempt to explain this effect has been reported
by Estreicher \cite{est99} employing the Hartree-Fock method and
H-terminated Si clusters. Accordingly, a Cu$_{\mathrm{i}}^{+}$ ion
bound either to a substitutional B, Al or Ga ion was moved to the
nearest $T$-site through the $H$ (hexagonal) ring, resulting in
metastable pairs with energies 0.45~eV, 0.64~eV and 0.68~eV above
that of their respective ground states. These figures and the respective
chemical trend were in good agreement with experimental observations,
but as carefully noted by the author \cite{est99}, they are not entirely
consistent with the plain-Coulomb model underlying Eq.~\ref{eq:effective_diffusivity}.
In particular, the results suggest that the dissociation process is
dominated by a short-range interaction, as the activation energy for
the very first step which breaks up the Cu$A$ pair exceeds the long-range
electrostatic dissociation energy.

The aim of this paper is two-fold: (i) to provide an in-depth analysis
of the dissociation process of the CuB pair in Si using first-principles
methods, and (ii) to inspect the interaction of CuB defects with photo-excited
carriers in the context of Cu-LID. Starting with the ground-state
structure of the CuB pair, the energy barriers for the first dissociating
steps are calculated. This allows us to inspect for any hypothetical
dissociation barrier greater than the sum of the long-range Coulomb
binding energy and the migration energy of Cu$_{\mathrm{i}}^{+}$.
Next, the stability of the intermediate metastable states of the CuB
pair is analysed as a function of the distance between $\mathrm{Cu_{i}^{+}}$
and $\mathrm{B_{s}^{-}}$, their electronic activity and interaction
with minority carriers (electrons) is calculated and discussed. Finally,
the energy to attain a fully dissociated pair is estimated by means
of independent supercells containing neutral Cu$_{\mathrm{i}}^{0}$
and B$_{\mathrm{s}}^{0}$ and subtracting the charge-transfer energy
due to electron donation from Cu to B.

\section{Method}

All calculations were computed using the $\mathtt{VASP}$ package
\cite{kre93,kre94,kre96,kre96a}, employing the Projector-Augment
Wave (PAW) \cite{blo94,kre99} method and a planewave basis set, within
the generalised gradient approximation to the exchange-correlation
potential among electrons, as proposed by Perdew, Burke and Ernzerhof
\cite{per96}. The PAW potentials used for $\mathrm{Cu}$, $\mathrm{B}$
and $\mathrm{Si}$ included the valence states given by the corresponding
electronic configurations: $3p^{6}3d^{10}4s^{1}$, $2s^{2}2p^{1}$
and $3s^{2}3p^{2}$. Following convergence tests, the maximum planewave
kinetic energy was set to $E_{\mathrm{cut}}=370$~eV.

Cu, B and CuB point defects were inserted into 216-atom Si supercells,
with a cubic shape and an optimised lattice constant $a=5.4687$~Å.
All defect structures were optimised using either a conjugate-gradient
method or a quasi-Newton algorithm, until the forces acting on the
atoms were converged within $1\times10^{-3}$~eV/Å. The self-consistent
electronic and magnetic relaxations were computed with an accuracy
of $1\times10^{-5}$~eV, and the band structures were sampled at
$\mathbf{k}$-points defined by $2\times2\times2$ Monkhorst-Pack
grids (avoiding the $\Gamma$-point) \cite{mon76}.

The marker method was employed to assess the electronic activity of
CuB complexes \cite{res99}. To this end, $\mathrm{Cu_{i}}(0/+)=E_{\text{C}}-0.15$~eV
\cite{ist97} and $\mathrm{B_{s}}(-/0)=E_{\text{V}}+0.045$~eV \cite{mor54}
were used as marker levels, corresponding to the calculated ionisation
potential and affinity energies $I\{\mathrm{Cu_{i}}(0/+)\}=6.247$~eV
and $A\{\mathrm{B_{s}}(-/0)\}=5.582$~eV, respectively. The spurious
long-range interactions arising from the periodic arrangement of localised
charged defects (immersed in a compensating background \emph{jellium})
was accounted for by an image-charge correction $E_{\mathrm{chg}}[D^{q}]$
as proposed in Refs.~\cite{fre09,kum14}. This term depends on the
shape of the ionised density of a particular defect $D$ in charge
state $q$, and was added to the total energies of defective supercells.
For the defects under scrutiny, and for first ionisation and affinity
energies, image-charge corrections were all close to $E_{\mathrm{chg}}\sim80$~meV.
Since the marker method relies on differences between the values of
$I$ (or $A$) calculated for the marker defect and the inspected
defect, these correction terms mostly cancel out, and the net contributions
are on the order of a few meV.

\begin{figure}[H]
\centering{}\includegraphics{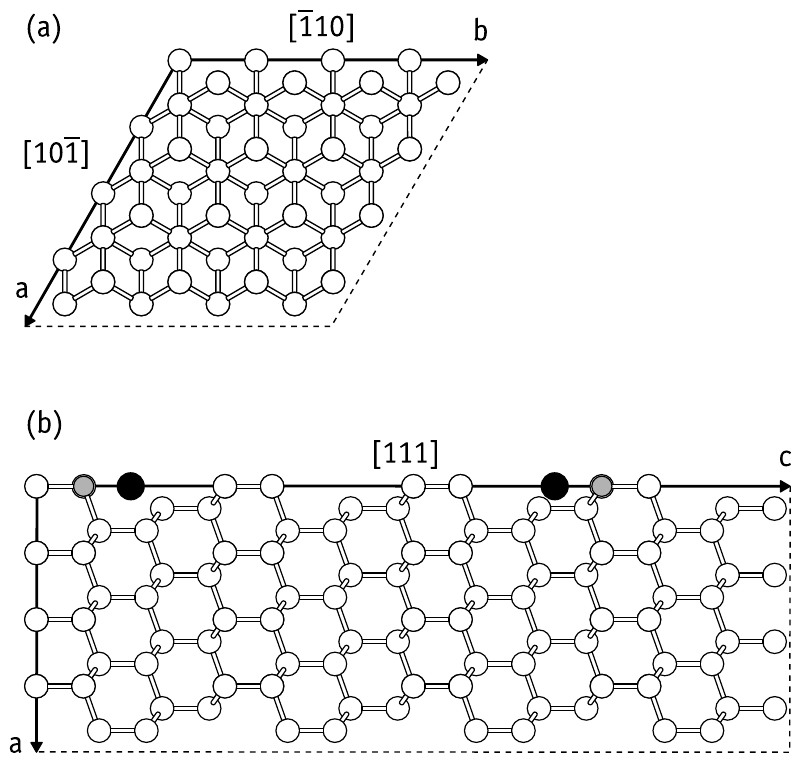}\caption{\label{fig:dipole}Projections of the hexagonal cell ($a=b=15.468$~Å,
$c=37.888$~Å) used in the calculation of the dipole-image interaction
energy, along (a) the $\text{[\ensuremath{\mathrm{111}}]}$ and (b)
$\text{[\ensuremath{\mathrm{\overline{1}10}}]}$ crystallographic
cubic directions, depicting two CuB defects with antiparallel dipole
moments. Si atoms are represented in white, B in grey and Cu in black.}
\end{figure}

The defects reported in this work comprise copper-boron pairs with
different separations. Each pair is expected to form an electric dipole,
not only within the hosting cubic cell, but also in all periodic cell
replicas. This introduces a spurious long-range dipole-image interaction
term in the total energy. This term is ill-defined with respect to
the origin of the coordinate system and it is not trivial to account
for \cite{mak95}. To estimate how large this dipole-image error can
be, we calculated the total energy of two CuB pairs (in their ground
state structure with $C_{3v}$ symmetry) in the same cell, which was
chosen to have a hexagonal shape (with 382 atoms and lattice parameters
$a=b=15.468$~Å, $c=37.888$~Å). Both defects were collinearly aligned
along the {[}111{]} crystallographic axis and separated by a distance
of $c/2$, with parallel dipole moments and antiparallel dipole moments
(as depicted in Fig.~\ref{fig:dipole}(b)), yielding total energies
denoted $E_{\uparrow\uparrow}$ and $E_{\uparrow\downarrow}$, respectively.
Since the latter energy corresponds to a system with zero net dipole
moment, we arrive at the dipole-image correction,

\begin{equation}
E_{\mathrm{dip}}=\left(E_{\uparrow\downarrow}-E_{\uparrow\uparrow}\right)/2=0.013\ \text{eV},
\end{equation}
per Cu-B pair. Such a small figure, allows us to safely neglect any
dipole-image corrections to the calculations presented below, even
considering error bars of the order of 0.1-0.2~eV. 

The potential energy surface (PES) governing the motion of a $\mathrm{Cu_{i}^{+}}$
ion close to $\mathrm{B_{s}^{-}}$ was investigated using the nudged
elastic band (NEB) method \cite{shep12}. Using this method, we were
able to calculate the energy barriers and the minimum energy paths
(MEP) connecting neighbouring tetrahedral interstitial sites using
7 intermediate structures. For the NEB calculations, the forces acting
on the atoms were converged within $1\times10^{-2}$~eV/Å.

\section{Results}

\begin{figure}[H]
\centering{}\includegraphics{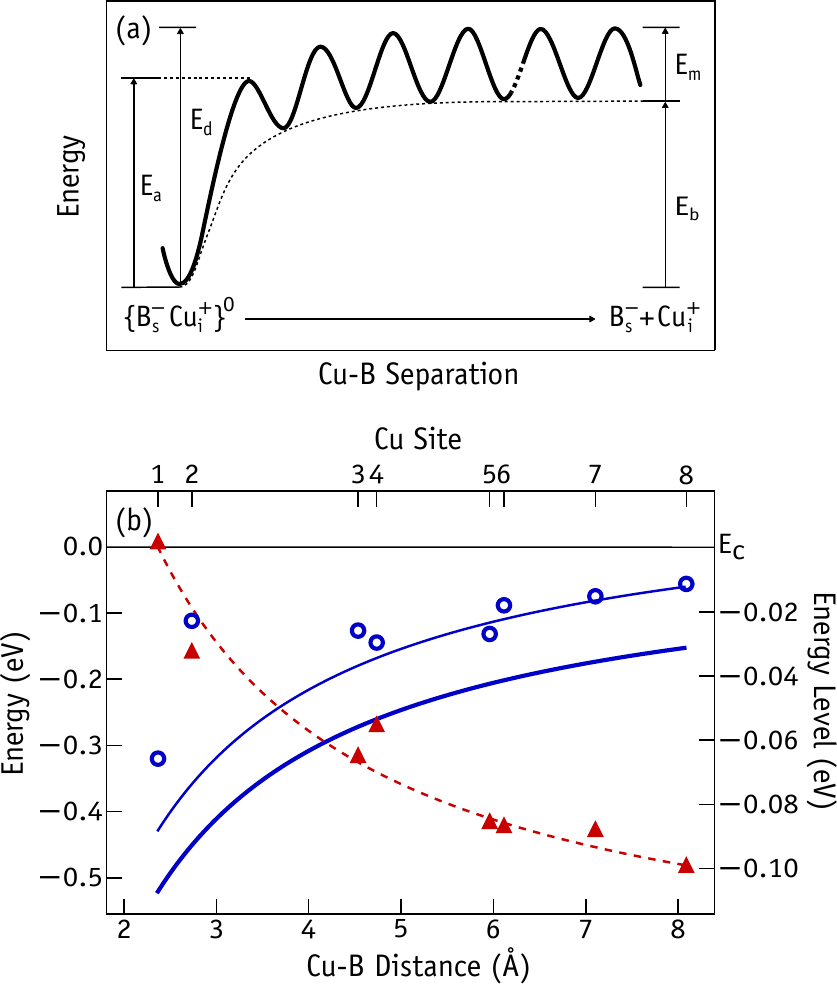}\caption{\label{fig:metastable_pairs}(a) Schematic representation of the potential
energy for the $\mathrm{\{Cu_{i}^{+}}\mathrm{B_{s}^{-}}\}^{0}$$\rightarrow\mathrm{Cu_{i}^{+}+\mathrm{B_{s}^{-}}}$
dissociation mechanism, as a function of the distance between $\mathrm{Cu_{i}^{+}}$
and $\mathrm{B_{s}^{-}}$ ions (see the beginning of this section
for a description of the energy parameters). (b) Relative energies
of CuB pairs (blue circles) and respective acceptor levels (red triangles).
Defect energies are given with respect to infinitely separated $\mathrm{Cu_{i}^{+}}$
and $\mathrm{B_{s}^{-}}$ ions (left-vertical axis). Electrical levels
are represented with respect to the bottom of the conduction band
($E_{\mathrm{C}}$) as indicated by the right-vertical axis. The lower-horizontal
axis indicates the distance between the ions, while the upper axis
refers the $n$-th neighbouring site of Cu with respect to B ($n=1,\cdots,8$).
The thick full line represents a screened Coulomb potential energy
$V_{\mathrm{C}}$ of two point charges of opposite sign, while the
thin full line shows $V_{\mathrm{C}}+0.1$~eV (see text). The dashed
line is a mere guideline.}
\end{figure}

The present description of the dissociation process of CuB pairs is
based on the first-principles calculation of four parameters, namely
(1) the activation energy ($E_{\mathrm{a}}$) required to break up
a stable CuB pair and transform it into a metastable configuration,
(2) the dissociation energy ($E_{\mathrm{d}}$) required to separate
Cu$_{\mathrm{i}}^{+}$ and B$_{\mathrm{s}}^{-}$ ions beyond their
long-range Coulomb interaction distance, (3) the migration energy
($E_{\mathrm{m}}$) of $\mathrm{Cu_{i}^{+}}$ in pristine crystalline
Si, and (4) the binding energy ($E_{\mathrm{b}}$) holding a CuB pair
together, with respect to uncorrelated $\mathrm{Cu_{i}^{+}}$ and
$\mathrm{B_{s}^{-}}$ defects. These parameters characterise the process,
in the manner depicted in Fig.~\ref{fig:metastable_pairs}(a).

\begin{figure}[H]
\noindent \centering{}\includegraphics{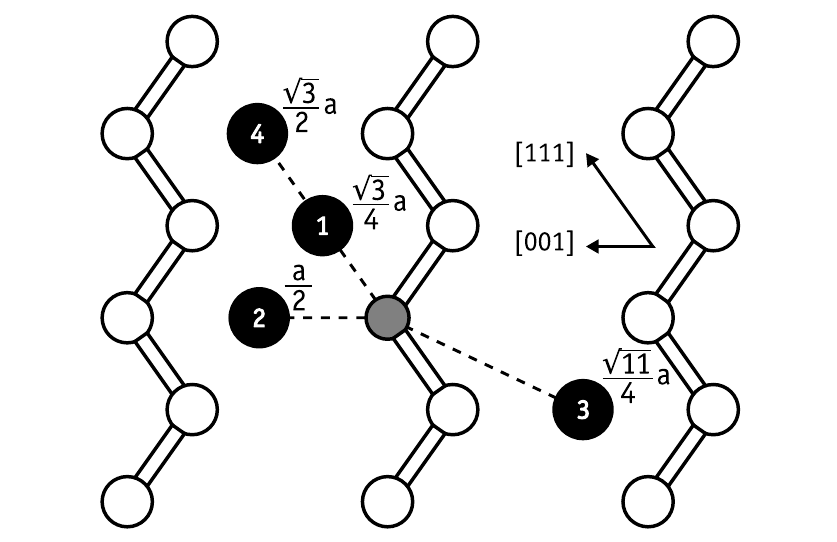}\caption{\label{fig:first_4_sites}Slice of a Si crystal along the $\mathrm{(1\bar{1}0)}$
plane containing a $\mathrm{B_{s}}$ dopant (grey atom close to the
centre) and its first four tetrahedral interstitial neighbouring sites,
which can be occupied by a $\mathrm{Cu}$ atom (numbered 1 to 4).
Site distances to the B atom are also quoted in units of the Si lattice
constant ($a$).}
\end{figure}

We start by reporting on the energetics of supercells containing the
CuB ground state and its first seven metastable configurations. The
energy of the pairs with respect to infinitely separated $\mathrm{Cu_{i}^{+}}$
and $\mathrm{B_{s}^{-}}$ defects are reported as blue circles in
Fig.~\ref{fig:metastable_pairs}(b). The calculation of the energy
reference (for the infinitely separated ions) is discussed in detail
further below. The left vertical axis in Fig.~\ref{fig:metastable_pairs}(b)
represents the reversed binding energy of a CuB complex for a Cu atom
sitting at the $n$-th neighbouring site relative to the B atom, as
indicated by the upper horizontal axis. The lower horizontal axis
represents the respective separation between Cu and B nuclei. The
first four configurations corresponding to the shortest Cu-B distances
are depicted in Fig.~\ref{fig:first_4_sites}. The energies of sites
2-8 are $0.21$, $0.19$, $0.18$, $0.19$, $0.23$, $0.25$ and $0.26\ \text{eV}$
above that of the ground state (site 1), respectively. These figures
reveal a prominent stabilisation of site 1 when compared to more distant
sites. Further, the interaction potential for sites 2-5 is approximately
constant. These results clearly indicate that the Cu-B interaction
involves a short-range effect which is not accounted for by a Coulomb
term alone.

In order to explore the PES of a moving Cu ion in the neighbourhood
of a B acceptor, we calculated the energy barriers $E_{\mathrm{a}}^{1,2}$
and $E_{\mathrm{a}}^{1,4}$ that account for Cu \emph{jumps} from
site 1 (ground state) to sites 2 and 4, respectively. Site 3 is too
distant (along the configurational space) from site 1, and is more
likely to be reached through site 2. For these calculations, the supercells
were kept neutral, and as we will see below, in the absence of above-bandgap
illumination the Cu atom hops around $\mathrm{B}_{\mathrm{s}}^{-}$
as a positively charged ion. The results are reported in Table~\ref{tab:assoc_dissoc}
below. Although $E_{\mathrm{a}}^{1,4}$ is $0.03\ \mathrm{eV}$ higher
than $E_{\mathrm{a}}^{1,2}$, the difference is within the precision
of the methodology, suggesting that the pair is able to break up either
along sites $1\text{-}4\text{-}\cdots$ or $1\text{-}2\text{-}3\text{-}\cdots$.
Regardless of the path, the very first activation barrier to the break-up
of a CuB pair is just over $E_{\mathrm{a}}\sim0.3$~eV. This is about
3 times the calculated migration barrier of isolated $\mathrm{Cu_{i}}^{+}$,
which was estimated as $E_{\mathrm{m}}=0.11$~eV using the NEB method
between neighbouring interstitial sites in a positively charged supercell
without boron (only 70~meV lower than the experimental value of 0.18~eV
for Cu migration in intrinsic Si without traps). Finally, considering
the relative energies of sites 2, 3 and 4, we also conclude that the
barrier for the reverse process, \emph{i.e.} for the capture of $\mathrm{Cu_{i}^{+}}$
by $\mathrm{B}_{\mathrm{s}}^{-}$, is about 0.1~eV and very close
to $E_{\mathrm{m}}$, suggesting that the short-range interaction
between Cu and B affects the association/dissociation kinetics at
the nearest sites only (perhaps 1-3). Beyond that, the Coulomb interaction
starts to dominate.

\begin{table}[H]
\centering{}\caption{\label{tab:assoc_dissoc}A comparison of the values calculated for
the relevant energy barriers in the Cu-B association/dissociation
process, compared to other calculations and experimental values in
the literature, in eV.}
\begin{tabular}{cccc}
\hline 
 & Calculated & Previous Calculations & Experimental\tabularnewline
\hline 
$E_{\text{m}}$ & 0.11 & 0.18 \cite{bac10}, 0.24 \cite{est99} & 0.18 \cite{ist98}\tabularnewline
$E_{\text{b}}$ & 0.32 & – & 0.43 \cite{note2}\tabularnewline
$E_{\text{d}}$ & 0.43 & $>0.69$ & 0.61 \cite{pre92}\tabularnewline
$E_{\text{a}}^{1,2}$ & 0.32 & – & –\tabularnewline
$E_{\text{a}}^{1,4}$ & 0.35 & 0.45 \cite{est99} & –\tabularnewline
\hline 
\end{tabular}
\end{table}

Next, the binding energy of the pair ($E_{\mathrm{b}}$) was calculated,
and the dissociation energy ($E_{\mathrm{d}}=E_{\mathrm{b}}+E_{\mathrm{m}}$)
was compared to $E_{\mathrm{a}}^{1,2}$ and $E_{\mathrm{a}}^{1,4}$.
This allowed us to investigate wether the macroscopic association/dissociation
kinetics is dominated by a short-range breaking/capture barrier, or
by the long-range shape of the Coulomb potential (added to the barrier
for migration of $\mathrm{Cu_{i}^{+}}$ through the lattice). The
value of $E_{\mathrm{b}}$ was obtained by considering the energy
gained by taking uncorrelated $\mathrm{Cu_{i}^{+}}$ and $\mathrm{B_{s}^{-}}$,
and combining them into a neutral $\{\mathrm{Cu_{i}^{+}B_{s}^{-}}\}^{0}$
pair. While the calculation of the final state is straightforward,
for the initial state the method must account for the compensation
of Cu by the B dopant. The energy required to infinitely separate
$\mathrm{Cu_{i}^{+}}$ from $\mathrm{B_{s}^{-}}$ was therefore calculated
as follows

\begin{equation}
E(\mathrm{Cu_{i}^{+}+B_{s}^{-}})=E(\mathrm{Cu_{i}^{0}})+E(\mathrm{B_{s}^{0}})-\Delta E_{\mathrm{CT}},\label{eq:binding}
\end{equation}
where $E(\mathrm{Cu_{i}^{0}})$ and $E(\mathrm{B_{s}^{0}})$ are the
energies of neutral $\mathrm{Cu_{i}}$ and $\mathrm{B_{s}}$ defects
calculated in separate supercells. The quantity $\Delta E_{\mathrm{CT}}$
accounts for the charge transfer energy that is released after demotion
of an electron from the Cu gap state to an infinitely distant B acceptor.
We obtain this quantity from $\Delta E_{\mathrm{CT}}=A\{\mathrm{B_{s}}(-/0)\}-I\{\mathrm{Cu_{i}}(0/+)\}=0.66$~eV,
which effectively is the energy difference between the calculated
$\mathrm{Cu_{i}}(0/+)$ and $\mathrm{B_{s}}(-/0)$ levels. We note
that the underestimated band gap of Si (due to the semi-local treatment
of the exchange-correlation energy), leads to analogous errors in
the calculation of the $\Delta E_{\mathrm{CT}}$ term and $E(\mathrm{Cu_{i}^{0}})+E(\mathrm{B_{s}^{0}})$
terms in Eq.~\ref{eq:binding}, which are expected to cancel out.
After fixing any stoichiometric mismatch in the number of Si atoms
in the calculation, the aforementioned considerations result in $E_{\mathrm{b}}=0.32$~eV,
which underestimates the experimental value by 0.1~eV only. We note
that if the charge transfer term had not been considered, the (first-principles)
binding energy for the $\mathrm{Cu_{i}^{0}}+\mathrm{B_{s}^{0}}\rightarrow\{\mathrm{Cu_{i}^{+}B_{s}^{-}}\}^{0}$
reaction would be 1.22~eV. Adding this value to the migration barrier
of copper yields a dissociation barrier of 1.4~eV, in obvious disagreement
with the observations. Nevertheless, our estimate for $E_{\mathrm{d}}$
is 0.43-0.50~eV, depending on whether we add the calculated or the
experimental value of $E_{\mathrm{m}}$ to the binding energy, and
this is well in line with a defect that is marginally stable at room
temperature. Finally, we note that $E_{\mathrm{d}}$ is larger than
both $E_{\mathrm{a}}^{1,2}$ and $E_{\mathrm{a}}^{1,4}$, indicating
that the dissociation of the pair involves the migration of $\mathrm{Cu_{i}^{+}}$
under the action of an attractive long-range Coulomb potential, which
it must eventually escape from.

Figure~\ref{fig:metastable_pairs}(b) shows a thick solid line that
represents the interatomic potential energy of two unitary point charges,
$V_{\mathrm{C}}=-\kappa_{\mathrm{e}}/r$, screened by the dielectric
constant of bulk Si ($\epsilon_{\mathrm{r}}=11.68$), where $k_{\mathrm{e}}$
is the Coulomb constant in eV\,Å. In the same figure we represent
$V(r)+0.1$~eV by a thin solid line. This allows us to directly compare
the first-principles data to the Coulomb potential, after adding 0.1
eV to the latter, the energy corresponding to the deviation between
the calculated and the experimental values of the binding energy.
From the figure, we conclude that, while a repulsive \emph{central-cell}
correction of about 0.1-0.2~eV has to be considered in order to obtain
agreement between the Coulomb model and the first-principles data
for sites 1 and 2, agreement is rather good for remote sites. When
compared to the Coulomb potential, the additional energy for sites
1 and 2 could be ascribed to a finite size effect like the increase
in the overlap of the 3d electrons of Cu to the excess of negative
charge close to the B$_{\mathrm{s}}^{-}$ ion.

The electronic activity of the various CuB pair structures was also
investigated. Despite the lack of evidence for carrier trapping or
recombination activity due to CuB complexes, the reason behind such
inert behaviour is actually unchartered. Further, this inactivity
has been asserted for the CuB ground state only, and nothing is known
about the metastable configurations. This aspect could be relevant
in the context of Cu-LID, particularly in p-type Si, since it could
support a direct connection between a hypothetical light-enhanced
dissociation of CuB pairs (and therefore the release of Cu) and the
observed decrease in minority carrier life-time.

The calculated ionisation energies of CuB pairs are mostly independent
of the distance between $\mathrm{B_{s}^{-}}$ and $\mathrm{Cu_{i}^{+}}$
ions, $I\{\text{CuB}(0/+)\}=5.50(5)$~eV (for sites 1-8). This figure
should be compared to $A\{\mathrm{B_{s}}(-/0)\}=5.582$~eV, indicating
that $\text{CuB}(0/+)$ transitions are about 80~meV below the boron
level, and effectively suggesting that none of the CuB defects have
donor levels in the gap. As a word of caution, we note that such small
energies are well within the error bar of the method, and an eventual
shallow hole trap related to $\text{CuB}(0/+)$ could be effectively
undetectable as it could lie just below the $\mathrm{B_{s}}(-/0)$
transition. We also point out that the weak dependence of the $\text{CuB}(0/+)$
levels on the Cu-B distance stems from the fact that they derive from
the shallow boron state (here fully occupied due to an electron donated
by Cu). The low amplitude and extensive spatial distribution of the
state means that it is unlikely to be sensitive to the relative position
of a localised $\mathrm{Cu_{i}^{+}}$ ion.

Next we investigated the acceptor activity of CuB pairs. Comparing
the affinities of the various CuB defects with the ionisation potential
of $\mathrm{Cu_{i}}$ we arrive at $\text{CuB}(-/0)$ levels ranging
from $E_{\mathrm{C}}-0.00$~eV (for the ground state) to approximately
$E_{\mathrm{C}}-0.10$~eV (for Cu at site 8). The results are reproduced
graphically in Figure~\ref{fig:metastable_pairs}(b) (refer to the
right-vertical axis), along with the dashed line that clearly indicates
a trend towards the isolated $\mathrm{Cu_{i}}(0/+)$ transition (experimentally
measured at $E_{\mathrm{C}}-0.15$~eV \cite{ist97}) when the ions
become infinitely separated. The calculations support the prevailing
ground-state model as an electronically inactive defect. However,
they also anticipate that the pairs turn into acceptors as soon as
the ion separation becomes larger than the first neighbouring distance.
All calculated $\text{CuB}(-/0)$ transitions edge the conduction
band minimum and derive from the $\mathrm{Cu_{i}}(0/+)$ transition,
which is shifted up in the energy scale due to Coulomb repulsion by
the $\mathrm{B_{s}^{-}}$ anion. An analogous argument was used to
explain the location of the $\mathrm{FeB}(-/0)$ level in Si as arising
from a $\mathrm{Fe_{i}}(0/+)$ transition displaced by $\mathrm{B_{s}^{-}}$
\cite{zha01}.

Since CuB pairs are marginally stable at room temperature, our results
suggest that in B-doped Si, photo-excited electrons could be trapped
at the Cu atom of a metastable CuB complex, effectively \emph{turning
off} the Coulomb binding potential between the pair. If this $\mathrm{B_{s}^{-}}\mathrm{Cu_{i}^{0}}$
state survives for a sufficiently long time before the electron is
emitted back into the conduction band, it is likely that a light-induced
dissociation enhancement of CuB pairs will take place, explaining
the doping/illumination dependence of Cu-LID on the basis of the change
in the dynamics of the binding/dissociation of mobile copper to/from
acceptors during a diffusion-limited process.

The nature of the defect behind Cu-LID remains unknown. Recently,
Lindroos and Savin \cite{lin16} hypothesised a possible involvement
of either substitutional copper or copper precipitates in the recombination-active
centre. It is also known that illumination leads to a decrease of
interstitial copper in silicon \cite{bel04}. Based on these premisses,
we investigated the light-induced transformation of $\mathrm{Cu_{i}}$
into $\mathrm{Cu_{s}}$ through the $\mathrm{Cu_{i}^{+}B_{s}^{-}}\stackrel{h\nu}{\rightarrow}\mathrm{Cu_{s}^{-}}+\mathrm{B_{i}^{+}}$
reaction, eventually made possible with the help of about 1~eV resulting
from the recombination between a photo-excited electron trapped at
a metastable CuB pair, and a free-hole. To this end we compared the
relative stability of $\mathrm{Cu_{i}B_{s}}$ and $\mathrm{Cu_{s}B_{i}}$
defects in neutral cells. Note that the first acceptor level of $\mathrm{Cu_{s}}$
(at $E_{\mathrm{V}}+0.09$~eV \cite{bro87}) lies below the donor
level of $\mathrm{B_{i}}$ (at $E_{\mathrm{C}}-0.13$~eV \cite{har82}),
implying that neutral $\mathrm{Cu_{s}B_{i}}$ is actually a $\mathrm{Cu_{s}^{-}B_{i}^{+}}$
ionic complex, where the boron atom is located close to the tetrahedral
interstitial site, toward the Cu site. We found that $\mathrm{Cu_{i}B_{s}}$
is more stable than $\mathrm{Cu_{s}B_{i}}$ by about 2.0~eV. It is
also reasonable to assume that the exchange transformation $\mathrm{Cu_{i}B_{s}}\rightarrow\mathrm{Cu_{s}B_{i}}$
should have a barrier in excess of the 2~eV energy difference. Although
we can not exclude the involvement of substitutional copper in the
Cu-LID defect, such a large barrier suggests that $\mathrm{Cu_{s}}$
cannot be formed directly from the interaction of $\mathrm{Cu_{i}}$
with $\mathrm{B_{s}}$, even with the assistance of above-bandgap
illumination.

\section{Conclusions}

In conclusion, a first-principles model of the association/dissociation
mechanism of CuB pairs in Si has been presented. This model is based
on calculations of (1) the activation energy ($E_{\mathrm{a}}$) required
to transform a stable CuB pair into a metastable configuration, (2)
the dissociation energy ($E_{\mathrm{d}}$) required to separate Cu
and B ions beyond their long-range Coulomb interaction distance, (3)
the migration energy ($E_{\mathrm{m}}$) of $\mathrm{Cu_{i}^{+}}$
in pristine crystalline Si, and (4) the binding energy ($E_{\mathrm{b}}$)
holding the CuB pair together. In calculating $E_{\mathrm{a}}$, it
was determined that the energy barrier that a Cu atom in the CuB ground
state complex has to overcome in order to reach the next interstitial
site is approximately $0.3\ \text{eV}$ smaller than the experimental
value of $E_{\mathrm{d}}\simeq E_{\mathrm{m}}+E_{\mathrm{b}}$, and
approximately $0.1\ \text{eV}$ smaller than the value of $E_{\mathrm{d}}$
calculated in the present work. This suggests that dissociation is
limited by the migration of $\mathrm{Cu_{i}^{+}}$ under the action
of an attractive long-range Coulomb potential (and not by a short-range
\emph{bond-breaking} barrier), and also supports the diffusion-limited
trapping model used to determine the experimental value of $E_{\mathrm{d}}$,
for large separations between Cu and B ions.

Further analysis based on the energetics of the stable CuB pair, and
of its 7 nearest-neighbouring metastable configurations, corroborates
previous suggestions that a short-range effect has an influence on
the stability of the ground state. In particular, we found that the
energies of the stable CuB pair, and of the first nearest-neighbouring
metastable configuration, are respectively $0.1\ \text{eV}$ and $0.2\ \text{eV}$
larger than expected for a Coulomb interaction alone. This is tentatively
assigned to the inability of the point-charge Coulomb model to describe
the overlap between the closed shell d-electrons of Cu$_{\mathrm{i}}^{+}$
and the excess of negative charge in B$_{\mathrm{s}}^{-}$. For Cu-B
separations greater than the 4th neighbouring distance, the relative
energies of the supercells are in relatively good agreement with a
long-range Coulomb potential mediated by the static dielectric constant
of Si.

The role of the association/dissociation dynamics of CuB pairs in
Cu-LID was investigated through calculations of the electronic activity
of all CuB configurations. The calculations suggest that while CuB
pairs do not have donor levels in the gap, all nearest-neighbouring
metastable configurations do have acceptor levels, ranging from $E_{\mathrm{C}}-0.03\ \text{eV}$
to $E_{\mathrm{C}}-0.10\ \text{eV}$, which become deeper with increasing
Cu-B separation, and tend to the $E_{\mathrm{C}}-0.15\ \text{eV}$
level of isolated $\mathrm{Cu_{i}^{+}}$. Notably, and given the marginal
stability of CuB pairs at room temperature, this electronic activity
suggests that under sun-light illumination a \emph{quasi-equilibrium}
population of photo-excited electrons may be trapped at Cu atoms in
metastable configurations of CuB pairs. Should the metastable configurations
last long enough, the trapped carriers effectively shield the Cu ions
from the attractive potential of $\mathrm{B}_{\mathrm{s}}^{-}$ until
they are emitted back into the conduction band. This explains the
dependence of the Cu-LID rate on doping/illumination in B-doped Si
on the basis of an increased release and diffusivity of Cu interstitials.

As the exact defect underlying Cu-LID remains unknown, and in an attempt
to conciliate the fact that (i) Cu-LID is reversed by forcing Cu out-diffusion
\cite{bou14} and that (ii) illumination leads to a decrease of interstitial
copper in Si \cite{bel04}, a possible light-induced transformation
of $\mathrm{Cu_{i}}$ into $\mathrm{Cu_{s}}$ through a $\mathrm{Cu_{i}^{+}B_{s}^{-}}\stackrel{h\nu}{\rightarrow}\mathrm{Cu_{s}^{-}}+\mathrm{B_{i}^{+}}$
reaction was also investigated. We found that the process is endothermic,
as the difference between the energies of two neutral supercells containing
$\mathrm{Cu_{i}^{+}B_{s}^{-}}$ and $\mathrm{Cu_{s}^{-}B_{i}^{+}}$
defects is $2.0\ \text{eV}$ (favouring the former structure). This
suggests that $\mathrm{Cu_{s}^{-}}$ cannot result from an interaction
between $\mathrm{Cu_{i}}$ and $\mathrm{B_{s}}$, even if assisted
by about $1\ \text{eV}$ resulting from an eventual non-radiative
recombination between a photo-excited electron trapped by a metastable
CuB pair and a free hole. 

\ack{}{}

This work was funded by the Fundação para a Ciência e a Tecnologia
(FCT) under projects PTDC/CTM-ENE/1973/2012 and UID/CTM/50025/2013,
and funded by FEDER funds through the COMPETE 2020 Program. The authors
would like to acknowledge the contribution of the COST Action MP1406.
Computer resources were provided by the Swedish National Infrastructure
for Computing (SNIC) at PDC.

\bibliographystyle{unsrt}
\bibliography{refs}

\begin{thebibliography}{10}
\bibitem{note2}
This figure was calculated by assuming that
  ${E}_{\mathrm{b}}={E}_{\mathrm{d}}-{E}_{\mathrm{m}}$.
\end{thebibliography}

\end{document}